\documentclass[fleqn,usenatbib]{mnras}

\usepackage{newtxtext,newtxmath}

\usepackage[T1]{fontenc}

\DeclareRobustCommand{\VAN}[3]{#2}
\let\VANthebibliography\thebibliography
\def\thebibliography{\DeclareRobustCommand{\VAN}[3]{##3}\VANthebibliography}

\usepackage{graphicx}	
\usepackage{amsmath}	
\usepackage{float}
\usepackage{ulem}
\usepackage{comment}

\graphicspath{{./}{figures/}}
\usepackage{xcolor}

\newcommand{\revised}[1]{{#1}}

\title[F\&P approach for EoR 21-cm signal]{Cosmic variance suppression in radiation-hydrodynamic modeling of the reionization-era 21-cm signal}

\author[A. Acharya et al.]{
Anshuman Acharya$^{1}$\thanks{E-mail: anshuman@mpa-garching.mpg.de},
Enrico Garaldi$^{1,2}$,
Benedetta Ciardi$^{1}$,
Qing-bo Ma$^{3,4}$       
\\
$^{1}$Max-Planck-Institut f\"ur Astrophysik, Garching 85748, Germany \\
$^{2}$ Institute for Fundamental Physics of the Universe, via Beirut 2, 34151 Trieste, Italy\\
$^{3}$School of Physics and Electronic Science, Guizhou Normal University, Guiyang 550001, PR China \\
$^{4}$Guizhou Provincial Key Laboratory of Radio Astronomy and Data Processing, Guizhou Normal University, Guiyang 550001, PR China
}

\date{Accepted XXX. Received YYY; in original form ZZZ}

\pubyear{2023}

\begin{document}
\label{firstpage}
\pagerange{\pageref{firstpage}--\pageref{lastpage}}
\maketitle

\begin{abstract}
The 21-cm line emitted by neutral hydrogen is the most promising probe of the Epoch of Reionization (EoR). Multiple radio interferometric instruments are on the cusp of detecting its power spectrum. It is therefore essential to deliver robust theoretical predictions, enabling sound inference of the coeval Universe properties. The nature of this signal traditionally required the modelling of $\mathcal{O}(10^{7-8} \, {\rm Mpc}^3)$ volumes to suppress the impact of cosmic variance. However, the recently-proposed Fixed \& Paired (F\&P) approach uses carefully-crafted simulation pairs to achieve equal results in smaller volumes. In this work, we thoroughly test the applicability of and improvement granted by this technique to different observables of the 21-cm signal from the EoR. We employ radiation-magneto-hydrodynamics simulations to ensure the most realistic physical description of this epoch, greatly improving over previous studies using a semi-numerical approach without accurate galaxy formation physics and radiative transfer. We estimate the statistical improvement granted by the F\&P technique on predictions of the skewness, power spectrum, bispectrum and ionized regions size distribution of the 21-cm signal at redshift $7 \leq z \leq 10$ \revised{(corresponding to $\geq80\%$ of the gas being neutral)}. We find that the effective volume of F\&P simulations is at least 3.5 times larger than traditional simulations. This directly translates into an equal improvement in the computational cost (in terms of time and memory). Finally, we confirm that a combination of different observables like skewness, power spectrum and bispectrum across different redshifts can be utilised to maximise the improvement.
\end{abstract}



\section{Introduction}\label{sec:intro}

One of the most important probes for studying the Epoch of Reionization (EoR; the period of the Universe when the first astrophysical objects formed and emitted UV photons that ionized the neutral hydrogen of the intergalactic medium) is the brightness fluctuation of the 21-cm line of neutral hydrogen as observed in emission or absorption against the cosmic microwave background radiation (CMB; \citealt{Field_1959,Hogan_1979,Madau_1997,Shaver_1999,Tozzi_2000,Ciardi_2003,Zaroubi_2013}). The 21-cm signal is a highly-forbidden hyperfine transition produced by the spin-flip transition of the ground state of neutral hydrogen. Thanks to the long lifetime of the excited state combined with the abundance of neutral hydrogen in the intergalactic medium (IGM) at the inception of the EoR, it can be used to trace the gas progressive ionization during the EoR. A statistical detection of the strength of the fluctuations of its brightness temperature can allow us to constrain our models of the early Universe and the formation of the first stars and galaxies. 
 
Multiple interferometric low-frequency radio telescopes have been designed to search for this signal, like LOFAR\footnote{Low-Frequency Array, \url{http://www.lofar.org}}, HERA\footnote{Hydrogen Epoch of Reionization Array, \url{https://reionization.org/}}, MWA\footnote{Murchison Widefield Array, \url{http://www.mwatelescope.org}}, PAPER\footnote{Precision Array to Probe EoR, \url{http://eor.berkeley.edu}} and the upcoming SKA\footnote{Square Kilometre Array, \url{https://www.skao.int/en}}. While the signal is yet to be detected, upper limits of the 21-cm power spectrum have been obtained \citep[for e.g.,][]{Mertens2020,HERA_2022,Trott_2020}. 
These have allowed to rule out some extreme astrophysical models by comparison with simulations \citep[e.g.,][]{Ghara2020,Mondal_2020,Greig_2021,Greig_2021lofar,Abdurashidova_2022}. 
Further efforts are being made to minimise observational errors by improving signal extraction techniques \citep[e.g.][]{Mertens_2023,Acharya_2023}. It is therefore imperative to concurrently improve numerical modeling, readying it for the detection of the 21cm signal. 
As of now, the main limitation is the trade-off between simulated volume and resolution achieved, forced by limits in computational power.

In fact, as the thermal noise is proportional to the wave-modes \citep{Koopmans_2015}, the best results from upcoming surveys with LOFAR, HERA, MWA and, subsequently, SKA are expected at the lowest $k$ values, i.e., $0.15 \leq k / (h~\rm cMpc^{-1}) \leq 1.5$, after sufficient mitigation of foreground contamination. This corresponds to large physical scales, and indeed e.g. \citet{Iliev_2014} found that simulations with box sizes greater than 100 $h^{-1}$ cMpc are necessary to accurately model the evolution and distribution of ionized regions. \revised{\citet{Kaur_2020}, instead, found that even larger boxes, i.e. $\gtrsim 175 h^{-1}$ cMpc, are necessary.} However, a simulation with a box size corresponding to $k \sim 0.15~h~\rm cMpc^{-1}$ is still limited by sample variance. 
The straightforward solution to this issue is to employ larger boxes in order to increase the sampling of the long wave modes observationally relevant. For example, in \cite{Ghara2020} a side length of $500~h^{-1}~\rm cMpc$ has been simulated, while the semi-numerical approach of \cite{Mesinger_2016} and \cite{Greig_2022} has modeled Gpc scales. Such large boxes also allow to study the topology of large ionized regions ($\gtrsim 10$ cMpc), which is especially important to understand the role of quasars. However, even these approaches have issues, as extremely large computational resources for running the simulations and storing the generated data are required, and running large number of such simulations would only be computationally affordable with the semi-numerical approaches. While they can be appropriate for some analyses, they also have their own limitations \citep[see for e.g.,][ for details]{Choudhury_2018}.

Additionally, there are other limitations on the extent of physics that can be explored with such simulations. For example, mini halos ($\lesssim 10^8~\rm M_{\rm \odot}$) are expected to play a significant role in reionization  \citep[see][]{Iliev_2005,Haiman_2001}, but their inclusion would either require a high enough mass resolution to be able to capture them in large boxes, or running significantly smaller boxes. The former exacerbates the computational requirements, while the latter limits the wave modes that can be sampled. 
A possible approach to tackle these problems is to use simulations with different box sizes, and combine their results. A recent example of this is an analysis of galaxy populations at $z \geq 8$ in \citet{Kannan_2023}, which, however, does not provide a truly combined picture, as the different boxes are manually re-scaled to account for resolution differences without actual convergence between simulations. There have also been suggestions to use techniques like deep learning to increase the resolution of large box simulations \citep[see][for applications to dark matter only $N$-body simulations]{Kodi_2020,Li_2021}, however these haven't been carefully tested for EoR redshifts.

Another approach is to run computationally-cheaper boxes but include the properties of the more-expensive, full-physics ones. Recently, \citet{Sinigaglia_2021} developed a method to map baryonic properties of the IGM onto DM-only simulations, but its applicability to the reionization epoch is unclear. 

An alternative approach is to simulate the full physics of reionization in small boxes that are augmented with techniques that compensate the limited volume. One way to do this is the Fixed and Paired (F\&P hereafter) approach described in \citet{Angulo_2016} and \citet{Pontzen_2016}. It has been shown to substantially reduce the cosmic variance in the matter power spectrum \citep[see][for examples]{Maion_2022,Klypin_2020,Villaescusa-Navarro_2018} with respect to the traditional approach for the same box size, bypassing the need to run a large number of smaller boxes. However, this reduction is substantial only for statistical properties (e.g., the stellar mass function), and not for local ones like the gas distribution of individual galaxies \citep[check][for more details]{Villaescusa-Navarro_2018}. The 21-cm signal is influenced by both the large-scale distribution of neutral hydrogen, and the local properties of the ionized regions around galaxies. Exploring the extent of improvement (if any) on the effective volume for estimating the summary statistics of the 21-cm signal using the F\&P approach can be helpful in minimising the required computational needs. This idea was explored in \citet{Giri_2023} by running a large number of realizations of the F\&P approach using {\tt 21cmFAST} \citep{Mesinger2007,Greig2015}. By comparing them with randomly generated simulations, they found that F\&P boxes could obtain the same precision in the 21-cm signal power spectrum as traditional boxes twice their size, allowing at least a factor of 4 reduction in computing costs. However, {\tt 21cmFAST} does not take into account baryonic hydrodynamics, thus implicitly assuming that baryons track dark matter. It also does not include a proper implementation of radiative transfer (RT) or galaxy properties, and thus becomes unreliable at scales below 1 cMpc. Hence it cannot provide an accurate picture of galaxy-scale effects on the 21-cm signal.

To increase confidence in the applicability of the F\&P method for improving EoR 21-cm signal studies, it is necessary to use a more realistic framework, i.e. one which includes baryonic hydrodynamics, RT and which models galaxy properties more accurately. For this, we employ a setup similar to the {\tt THESAN} simulations \citep{Kannan_2022, Garaldi_2022,Smith_2022,Garaldi_2023}, detailed in Section~\ref{sec:methods}. We discuss the results of various summary statistics in Section~\ref{sec:results}, and the improvement on using the F\&P approach in Section~\ref{sec:discuss}, while we give our conclusions in Section~\ref{sec:summary}.

\section{Methodology}\label{sec:methods}

\subsection{The 21-cm signal}

The brightness temperature fluctuations of the 21-cm signal are given relative to the CMB temperature for any patch of the IGM as \citep[see][]{Furlanetto_2006eq}: \begin{multline}\label{eq:dtb} \delta T_{ \rm b} = 27 x_{\rm HI} (1 + \delta_{\rm B}) \left(1 - \frac{T_{\rm CMB}}{T_{S}} \right) \\
        \times\biggl[ \left( \frac{\Omega_{\rm B}h^2}{0.023} \right) \left( \frac{0.15}{\Omega_{\rm m} h^2} \frac{1+z}{10} \right)^{1/2} \biggr]~\rm mK
\end{multline} where $x_{\rm HI}$ is the fraction of neutral hydrogen, $\delta_{\rm B}$ is the fractional overdensity of baryons, $T_{\rm S}$ is the hydrogen spin temperature, $T_{\rm CMB}$ is the temperature of the CMB photons at redshift $z$, $\Omega_{\rm m}$ is the total matter density, $\Omega_{\rm B}$ is the baryon density, and $h$ is the Hubble constant in units of 100 kms$^{-1}$cMpc$^{-1}$. \revised{Above we assume that the spin temperature is coupled to the gas temperature, i.e., $T_{\rm S} = T_{\rm gas}$, where $T_{\rm gas}$ is the gas temperature self-consistently calculated in the simulations.}

In the next section, we introduce the simulation used to generate mock differential brightness temperature ($\delta T_{\rm b}$) maps.

\subsection{Simulations}

Our setup is inspired by the {\tt THESAN} simulations. We run a suite of radiation-magneto-hydrodynamic simulations that utilize the moving-mesh hydrodynamics code {\tt AREPO} \citep{Springel_2010,Weinberger_2020}, which includes a gravity solver based on the hybrid Tree-PM method \citep{Barnes_1986}, a quasi-Lagrangian Godunov method \citep{Godunov_1959} based hydrodynamics solver implemented on an unstructured Voronoi mesh grid \citep{Vogelsberger_2020} and the radiative transfer extension {\tt AREPO-RT} \citep{Kannan_2019} for a self-consistent treatment of ionizing radiation. We include the production and propagation of ionizing photons in three energy bins relevant for hydrogen and helium photoionization ([13.6, 24.6, 54.4, $\infty$] eV). Further, we emply a non-equilibrium thermochemistry solver to model the coupling of radiation fields to gas. For the luminosity and spectral energy density of stars, we use a complex function of age and metallicity calculated using the Binary Population and Spectral Synthesis models \citep[BPASS v2.2.1;][]{Eldridge_2017}, modeling the unresolved birth cloud with a uniform escape fraction of $f_{\rm esc}$ = 0.37. We note that we do not perform a recalibration of this parameter with respect to {\tt THESAN} since (i) it mostly impacts the final phases of reionization, while we are interested in the initial ones, and (ii) our goal is to compare methods for initial condition generation, so a slightly-inaccurate reionization history is not expected to affect at all our results. For further details on the {\tt THESAN} simulations, see \citet{Kannan_2019,Kannan_2022}. 

All our simulations have a box-size of $L = 95.5~\rm cMpc$, and $N = 2 \times 525^3$ particles, giving a dark matter and baryonic particle mass of $m_{\rm DM} = 2.0 \times 10^8~\rm M_{\odot}$ and  $m_{\rm gas} = 4.7 \times 10^6~\rm M_{\odot}$, respectively. The gravitational softening length for the star and dark matter particles is set to 6.0 ckpc, which is also the minimum value for the adaptively softened gas cells according to cell radius. The cosmological parameters are taken from \citet{Planck_2016} as $h = 0.6774$, $\Omega_{\rm m}=0.3089$, $\Omega_{\Lambda}=0.6911$, $\Omega_{\rm b}=0.0486$, $\sigma_{\rm 8}=0.8159$ and $n_{s}=0.9667$. In the initial conditions the gas is assumed to follow the DM distribution perfectly, with a primordial hydrogen and helium fractions of $X = 0.76$ and $Y = 0.24$, respectively. We start the simulations from $z_{\rm ini} = 49$, generating 54 snapshots between $z = 20~\rm to~7$. 

\revised{Due to the chosen particle number, our dark matter halo masses are $\gtrsim 10^{10}~\rm M_{\odot}$, leading to reionization being driven by relatively massive galaxies. The resulting topology of reionization therefore somewhat differs at small scales from the one that would be produced including lower mass, more abundant, galaxies. This will impact the 21-cm signal and slow down the reionization process \citep[which is driven by $M_\mathrm{star} \sim 10^7 \, \mathrm{M}_\odot$ galaxies as shown in][]{Rosdahl_2022, Yeh_2023, Kostyuk_2023}. However, these effects will equally affect all simulations, so will be factored out by our comparative analysis. A higher mass resolution would be immensely computationally expensive to run the number of simulations necessary to perform the statistical analyses discussed in subsequent sections of this work. As discussed earlier, the necessity of simulating large physical scales in order to compare with upcoming surveys prevents us from employing smaller (and thus computationally cheaper) boxes.}

\revised{Once a significant fraction of neutral hydrogen is reionized, the improvement of using the F\&P approach to study the 21-cm signal is expected to saturate. This is depicted in Figures~\ref{fig:ps_improve} and~\ref{fig:Bequi_improve}, showing the improvement factor calculated according to the methodology of Section~\ref{sec:improve}. Thus, we only run the simulations down to $z_{\rm fin} = 7$, which corresponds to $\langle x_{\rm HI} \rangle \approx 0.8$.}

Lastly, as done in \citet{Kannan_2022}, we also save Cartesian data output at a higher cadence (243 outputs between $z = 16$ to $z = 7$) by gridding the simulation data onto a regular Cartesian grid
employing a first order Particle-In-Cell approach. We use a $256^3$ grid, i.e. each cell is $\sim 372$ ckpc in size.

\begin{figure}
\centering
\includegraphics[width=1.0\columnwidth,keepaspectratio]{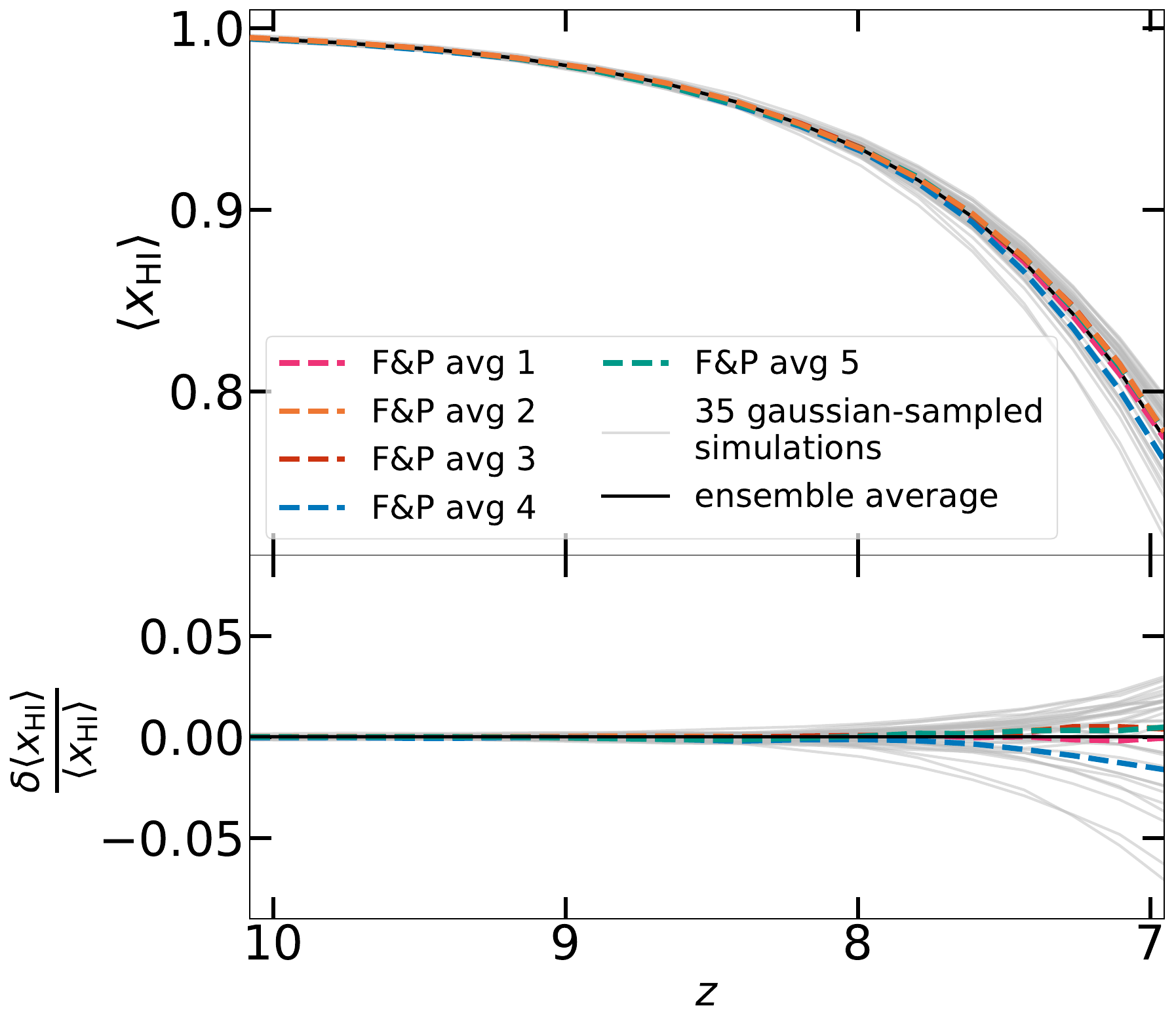} \\
\caption{Evolution of the volume averaged neutral hydrogen fraction versus redshift for GIC simulations (grey solid), their ensemble average (black solid) and five F\&P averages (orange, magenta, red, blue and green dashed). We note that at $z$=7 $\langle x_{\rm HI} \rangle$ ranges between 0.72 and 0.82 for the GIC simulations.}  
\label{fig:xHI_vs_z}
\end{figure}

\subsubsection{The Fixed \& Paired approach}\label{sec:fandp}

\begin{figure*}
\centering
\includegraphics[width=2.0\columnwidth,keepaspectratio]{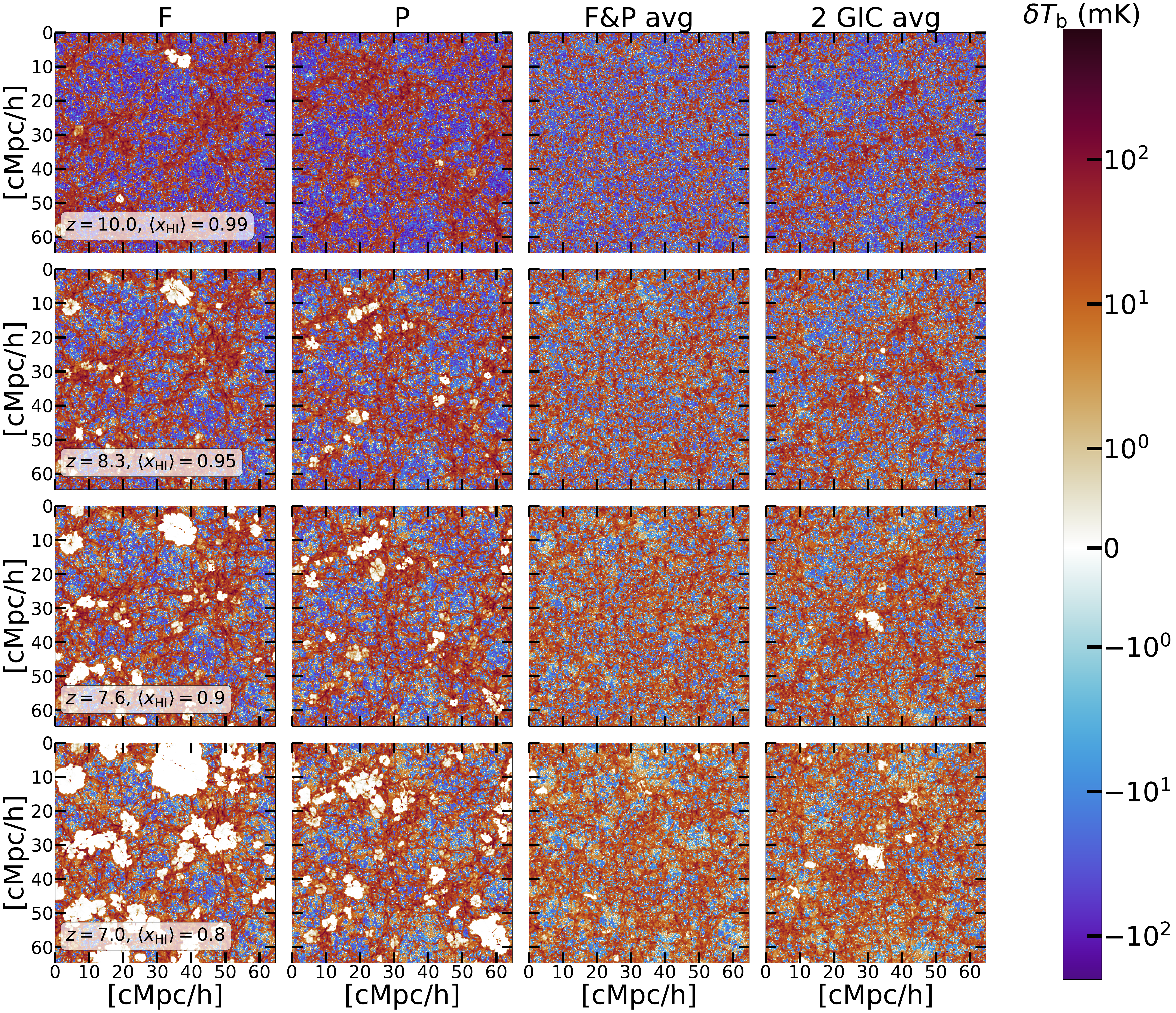}
\caption{Middle slices of the $\delta T_{\rm b}$ maps of (from left to right): one of the fixed simulations, its corresponding pair, their average, and the average of 2 GIC simulations, at $z = 10, 8.3, 7.6, 7$ (with $\langle x_{\rm HI} \rangle = 0.99, 0.95, 0.90, 0.80$ respectively). 
The average of 2 GIC simulations shows clear regions of high (and low) $\delta T_{\rm b}$, unlike the F\&P average. \revised{With decreasing redshift, such regions begin to form for the F\&P average as well, but a closer analysis of the $\delta T_{\rm b}$ summary statistics is necessary to analyse the difference from GIC simulations.}
}  
\label{fig:dTb_slice}
\end{figure*}

The F\&P approach was first proposed by \citet{Angulo_2016} and is based on the creation of a special pair of initial conditions (ICs) that, when employed together, significantly reduce the impact of cosmic variance. 
In the traditional approach to ICs creation, an initially uniform and isotropic distribution of tracers particles is perturbed \citep[following e.g.][]{Zeldovich_1970} to induce density perturbations: \begin{equation}
    \delta (\mathbf{k},z_{\rm ini}) = \sqrt{P(\mathbf{k},z_{\rm ini})} e^{i \theta_{\mathbf{k}}}
\end{equation} where the phase $\theta_{\mathbf{k}}$ is sampled from a flat distribution in the range $[0, 2\pi]$, and the power spectrum amplitude $P(\mathbf{k},z_{\rm ini})$ is sampled from a Gaussian distribution centered on its expectation value $E[P(\mathbf{\hat{k}},z_{\rm ini})]$ (necessary to produce a Gaussian random field).

In the F\&P approach, the power spectrum modes are fixed to their expectation values. This produces the \textit{fixed} initial conditions, while the \textit{paired} one is obtained by reversing the phase associated to each particle displacement. By combining simulations run with these two sets of ICs, the variance-induced fluctuations (which in traditional ICs mainly affect the large-scale modes, where the sampling of the power spectrum amplitude is scarce and therefore more susceptible to deviations from its expectation value) are suppressed up to the fourth perturbative order \citep{Angulo_2016}. Additionally, despite breaking the Gaussianity of the generated field, this approach does not induce unwanted features, as shown by e.g., \citet{Angulo_2016} and \citet{Chartier_2021} for the matter power spectrum, bispectrum and the halo mass function, by \citet{Anderson_2019} for Lyman-$\alpha$ forest power spectra, and for a variety of other quantities by \citet{Villaescusa-Navarro_2018} and \citet{Klypin_2020}.

In this work, we generate 5 such pairs of F\&P simulations, in order to explore the effects of the random sampling of density perturbations. \revised{To minimise the effect of randomness, for each pair we fix the seed for the random number generator used for stochastic algorithms like the star formation prescription. Further, we run all the simulations described here on the same machine with the same hardware configurations. As {\tt AREPO} is coded to be binary identical in such conditions, it allows us to avoid floating point errors building up and biasing our results.}

We compare the averages of these 5 \revised{F\&P} pairs against 35 traditional Gaussian-sampled initial conditions based simulations (hereafter referred to as GIC). As an example,  in Figure~\ref{fig:xHI_vs_z} we show the reionization histories of the 5 pairs (magenta, orange, red, blue, green dashed), as well as that of the GIC simulations (grey solid). From the figure, we note that the reionization histories of the F\&P averages cluster close to the average of those of the GIC simulations. This is a consequence of the fact that the F\&P method ensures that the F\&P averages closely match the halo mass function, while GIC simulations can spuriously have an excess/dearth of very bright sources. 

In Figure~\ref{fig:dTb_slice}, we show maps of  $\delta T_{\rm b}$ at different redshifts ($z = 10, 8.3, 7.6, 7$; these correspond to $\langle x_{\rm HI} \rangle = 0.99, 0.95, 0.9, 0.8$) for one of the fixed simulations, its corresponding pair, their average, and the average of two randomly chosen GIC simulations. \revised{We choose two GIC simulations at random to provide a visual comparison of results obtained from averaging them as opposed to averaging an F\&P pair.} Note that due to the phase inversion used for generating the initial conditions of the pair, the regions of high $\delta T_{\rm b}$ in the fixed simulation roughly overlaps with regions of low $\delta T_{\rm b}$ in the pair. This is evident in their average, which does not have specific regions of high (or low) $\delta T_{\rm b}$, unlike the average of the two GIC simulations. \revised{With decreasing redshift, we see such regions beginning to form for the F\&P average as well, but a closer analysis of the $\delta T_{\rm b}$ summary statistics is necessary to analyse the difference from GIC simulations.}

In the next section, we compare the ensemble average of the GIC simulations (taken to be the ``true" value) against the F\&P average simulations. We show qualitative comparisons between two of the F\&P averages and the true value in Section~\ref{sec:results}, and a more quantitative analysis using all five of the averaged simulations in Section~\ref{sec:discuss}.

\section{Analysis and Results}\label{sec:results}

In this section, we compare the various summary statistics of the 21-cm signal for the GIC simulations and their ensemble average versus the F\&P averages. In particular, we focus on the skewness, the power spectrum, and the bispectrum. In Sections~\ref{sec:skew},~\ref{sec:ps} and~\ref{sec:bs} we analyse their behaviour at various redshifts, focusing on $z = 10, 9, 8,~\rm and~7$.  We also compare the ionized bubble size distribution across these redshifts in Section~\ref{sec:bubble}.

\subsection{Skewness}\label{sec:skew}

The skewness ($\Tilde{\mu}_{\rm 3,b}$) is a statistical measure of asymmetry in a distribution, i.e., as the name suggests, it quantifies how skewed a distribution is around its mean value. $\Tilde{\mu}$ = 0 indicates a symmetric distribution. Here,  we use the definition of skewness from \citet{Ma_2021}, given as: \begin{equation} \label{eq:skewness}
    \Tilde{\mu}_{\rm 3,b} = \frac{\mu_3 (\delta T_{\rm b})}{\mu_2 (\delta T_{\rm b})^{3/2}} = \rm E \Bigg[\frac{(\delta T_{\rm b} - \langle \delta T_{\rm b} \rangle)^3}{\sigma_{\delta T_{\rm b}}^3} \Bigg]
\end{equation} where $\mu_i$ is the $i$-th central moment, and $\rm E[]$ is the expectation value at each snapshot where we calculate the skewness. In our case, this would be the volume average. 

\begin{figure}
\centering
\includegraphics[width=1.0\columnwidth,keepaspectratio]{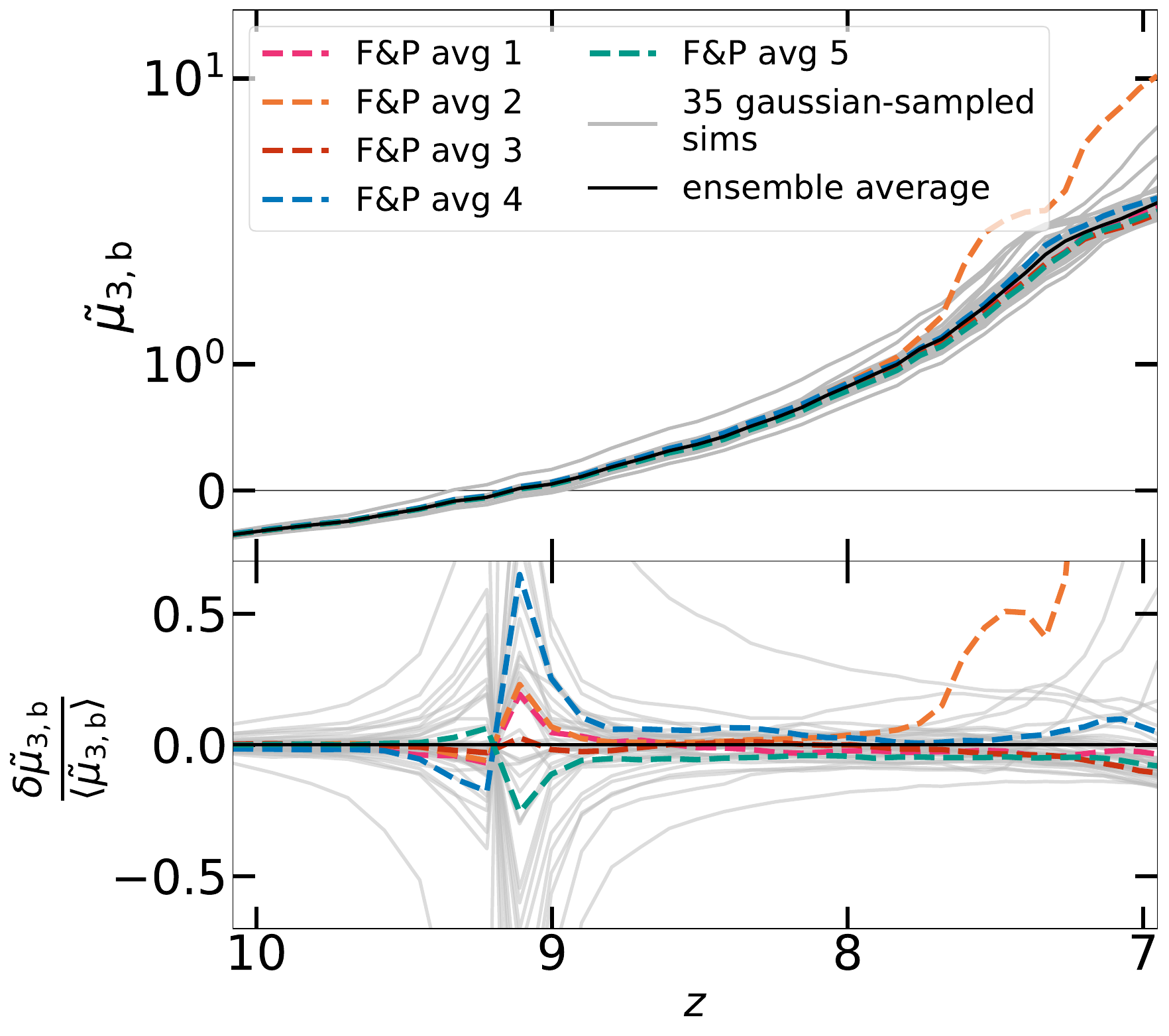}
\caption{Redshift evolution of skewness for 35 GIC simulations (grey solid), their ensemble average (black solid), and the five F\&P averages (orange, magenta, red, blue and green; dashed). 
}
\label{fig:skew}
\end{figure}

In Figure~\ref{fig:skew}, we compare the evolution of skewness versus redshift from $z= 10~\rm to~7$ for the 35 GIC simulations, their ensemble average, and the 5 F\&P average simulations. We note that the F\&P averages are a good estimate for the true value of the skewness in most cases. For the 2nd F\&P average, the deviation is larger at $z \lesssim 8$, which is indicative of more inhomogeneity in the $\delta T_{\rm b}$. We find that this anomalous skewness is caused by the fixed simulation in the second pair having a chance association of galaxies with strong black hole feedback and high output of ionizing photons. While such cases are rare, they are physically possible and highlight how the relevance of galactic processes in the production of the 21-cm signal hinders the improvements granted by the F\&P approach. We also note that such cases are not completely captured by e.g. the \texttt{21cmFAST} simulation used in \citet{Giri_2023}. 

\revised{This anomalous skewness however, is not expected to cause a major effect on other summary statistics like the power spectrum and bispectrum, because the effect is localised around few simulated galaxies. Thus, not only the scales involved are significantly smaller than the scales we consider in the next sections, but this effect is driven by a statistically-insignificant number of objects. To ensure this, we checked that by replacing in Equation~\ref{eq:skewness} the volume average with the median, the anomalous behavior vanishes. 
To make this more evident, in all following figures concerning summary statistics ($0.15 \leq k / (h~\rm cMpc^{-1}) \leq 2.0$) we elected to show the F\&P pair that most closely follows the ensemble-averaged skewness and the pair that most deviates from it. There is no significant difference between these two.} 

\subsection{Power Spectrum}\label{sec:ps}

\begin{figure}
\centering
\includegraphics[width=0.99\columnwidth,keepaspectratio]{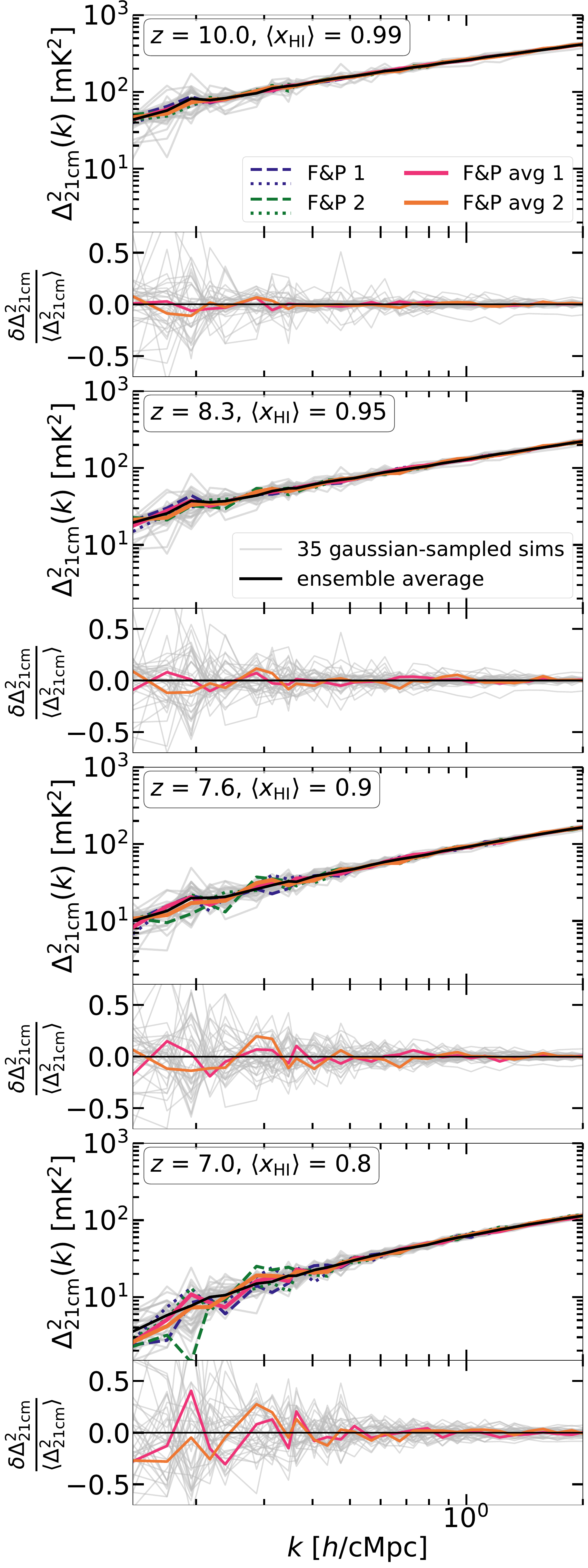}
\caption{ \textbf{Top:} Power spectra of GIC simulations (grey solid), their ensemble average (black solid), two F\&P pairs (purple and green, dashed and dotted) and their averages (magenta and orange, solid) for $z = 10, 8.3, 7.6,~\rm and~7$ from left to right. \textbf{Bottom:} The normalised deviation of each random simulation and F\&P average from the ensemble average.}
\label{fig:powerspectra}
\end{figure}

The power spectrum is expected to be the first detectable statistic of the 21-cm signal during the EoR, and thus is of particular interest. So analysing the improvement in modelling, if any, with the F\&P method is important. The power spectrum of $\delta T_{\rm b}$ as defined in Equation~\ref{eq:dtb} is given by: \begin{equation}\label{eq:ps}
P_{\rm 21cm} (\mathbf{k}) = \delta_D(\mathbf{k}+\mathbf{k}') \langle \delta T_{\rm b}(\mathbf{k}) \delta T_{\rm b}(\mathbf{k}') \rangle
\end{equation} where $\delta_D$ is the Dirac function, $\delta T_{\rm b}(\mathbf{k})$ is the differential brightness temperature in Fourier space, and $\langle ... \rangle$ is the ensemble average. In this work, we use the normalized form of the power spectrum given by:\begin{equation}\label{eq:psnorm}
    \Delta^{2}_{\rm 21cm} = \frac{k^3}{2\pi^2} \times P_{\rm 21cm}.
\end{equation} 

\begin{figure*}
\centering
\includegraphics[width=2.0\columnwidth,keepaspectratio]{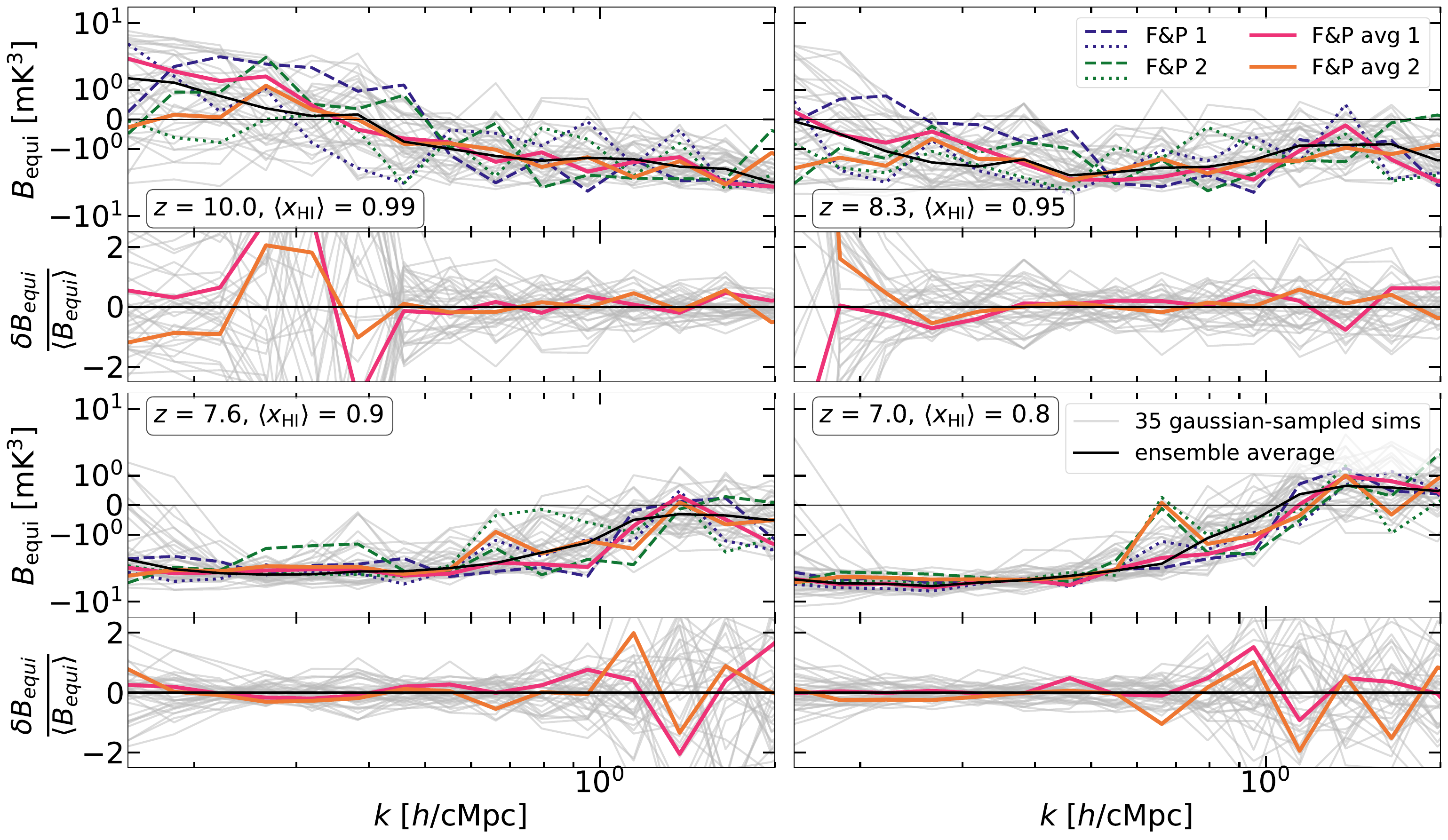}
\caption{ \textbf{Top panels:} Equilateral triangle bispectra ($B_{\rm equi}$)  for $0.15 \leq k / (h~\rm cMpc^{-1}) \leq 2.0$ of GIC simulations (grey solid), their ensemble average (black solid), two F\&P pairs (purple and green, dashed and dotted) and their averages (magenta and orange, solid) for $z = 10, 8.3, 7.6,~\rm and~7$, clockwise from upper left. \textbf{Bottom panels:} The normalised deviation of each random simulation and F\&P average from the ensemble average. 
}  
\label{fig:equi_bispectra}
\end{figure*}

In Figure~\ref{fig:powerspectra} we show the power spectra of the GIC simulations (in grey), their ensemble average (in black), and the F\&P averages (magenta and orange) for 2 pairs (purple and green, with the fixed simulation in dashed, and their corresponding pairs in dotted lines) out of the 5 generated in Section~\ref{sec:fandp}. While we have analysed results for all of the five generated F\&P pairs 
(see Section~\ref{sec:improve} for a quantitative comparison), for the sake of visual clarity we show only two out of the five pairs. \revised{We explicitly choose the second F\&P pair average (orange) to check if its deviation in skewness leads to any difference in behaviour as compared to one of the other four pairs (magenta).}

\revised{We note that the F\&P averages match the ensemble average across all redshifts. While their deviation from the ensemble average increases with decreasing redshift, it is still less than that of the individual GIC simulations (see lower panels of Figure~\ref{fig:powerspectra}).}
This is understandable as a combination of two effects. First, the F\&P approach is expected to yield improvements on the predictions of features that are primarily governed by the large-scale structure \revised{(as they are tied to the matter power spectrum)}, while it does not provide a statistical improvement on the prediction of galaxy properties \citep[][]{Villaescusa-Navarro_2018}, such as ionizing photons output, as they are entirely dominated by local physics. Secondly, the period of emergence of ionized regions is one where the 21-cm signal becomes increasingly non-Gaussian. 
Thus, the information content in the power spectrum is reduced in this period as compared to earlier and later redshifts, which have a homogenized distribution of neutral and ionized hydrogen, respectively. 
\revised{While the contribution from the large-scale structure remains the same, the fluctuations of $\delta T_{\rm b}$ are increasingly affected by the presence of ionized gas, which in our simulation is dominated by large and isolated ionized regions. While this mostly affects the power spectrum at small scales, the cumulative fluctuations over large scales can also show up. Thus for the F\&P method we observe deviations in the 21-cm signal power spectrum which are larger than those in the matter power spectrum as redshift decreases. However, we note that these deviations are still smaller than those of the individual GIC simulations. Nevertheless, it is necessary to check if an F\&P pair average power spectrum is more likely to minimize cosmic variance as opposed to an average of two random GIC simulations. Thus, we quantify the improvement on using the F\&P method in Section~\ref{sec:improve}, where we consider all our F\&P pair averages.}

While discrepancies between our results and \citet{Giri_2023} could possibly be due to the difference in methodologies adopted for comparing the F\&P method versus traditional generation of initial conditions, contributions from better handling of galaxy-driven physics in our simulations are also possible. It is however difficult to discern the extent of the effect galaxy-driven physics has on our results. \revised{Finally, it is also possible that our ensemble average is biased because it is an average of just 35 simulations. Using significantly larger number of GIC simulations may rule out this issue. However, we note no appreciable difference in the ensemble average once more than $20$ GIC simulations have been used, and thus refrain from running additional simulations.}

\subsection{Bispectrum}\label{sec:bs}

\begin{figure*}
\centering
\includegraphics[width=2.0\columnwidth,keepaspectratio]{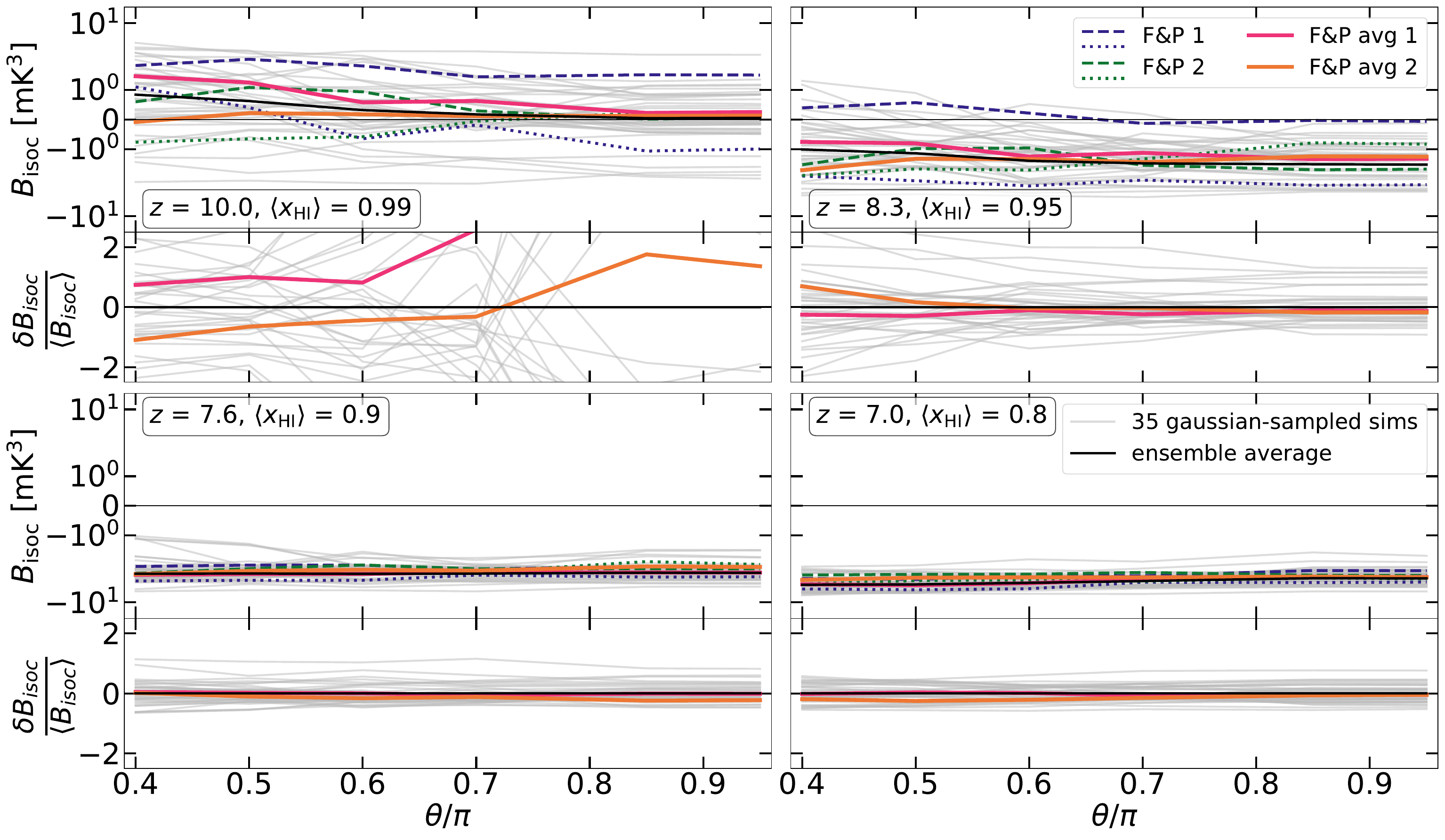}
\caption{ \textbf{Top panels:} Isosceles triangle bispectra ($B_{\rm isoc}$)  for $k_1 = k_2 = 0.2~h~\rm cMpc^{-1}$ for $z = 10, 8.3, 7.6,~\rm and~7$, clockwise from upper left. Colours and linestyles follow Figure~\ref{fig:equi_bispectra}. \textbf{Bottom panels:} The normalised deviation of each random simulation and F\&P average from the ensemble average. 
}  
\label{fig:isoc_bispectra}
\end{figure*}

As discussed in the previous section, the non-Gaussianity of the $\delta T_{\rm b}$ increases with the growth of ionized regions as redshift decreases. This means that a statistic that focuses only on the Gaussian parts of the signal, like the power spectrum, encapsulates less and less information as we move to lower redshift, into the regime of $\langle x_{\rm HI} \rangle \lesssim 0.9$. Therefore, we turn to higher-order statistics to assess whether the F\&P average is still a good approximation of the ensemble average of simulations in this regime, since they are able to capture the non-Gaussian features of the $\delta T_{\rm b}$. While skewness as discussed in Section~\ref{sec:skew} would show some broad non-Gaussian features of the signal, it is still a one point
statistic, i.e. it will not quantify the correlation of the signal
between different Fourier modes. Thus, we now focus on the bispectrum as defined in \citet{Majumdar_2018}:

\begin{equation}
    b_{\rm 21cm} (\mathbf{k_1},\mathbf{k_2},\mathbf{k_3}) = \delta_D (\mathbf{k_1} + \mathbf{k_2} + \mathbf{k_3}) \langle \delta T_{\rm b}(\mathbf{k_1}) \delta T_{\rm b}(\mathbf{k_2}) \delta T_{\rm b}(\mathbf{k_3}) \rangle
\end{equation} where $\mathbf{k_1},\mathbf{k_2},\mathbf{k_3}$ are the Fourier space wave numbers, $\delta_D$ is the Dirac delta function and $\langle \delta T_{\rm b}(\mathbf{k_1}) \delta T_{\rm b}(\mathbf{k_2}) \delta T_{\rm b}(\mathbf{k_3}) \rangle$ is a measure of the number of triangles (weighted by the $\delta T_{\rm b}$ values at their vertices) of different configurations formed by wave numbers $\mathbf{k_1},\mathbf{k_2}~\rm and~\mathbf{k_3}$. The different triangles can be formed by varying the magnitude of the wave numbers. 

To evaluate the bispectra from our simulations, we use the publicly available {\tt BiFFT} package \citep{Watkinson_2017}, which employs a  Fourier transform based technique \citep[as described in ][]{Scoccimarro_2015,Sefusatti_2016} much faster rather than the more traditional approach of counting individual triangles, while still providing consistent results. 
Following \citet{Majumdar_2020}, we normalize $b_{\rm 21cm} (\mathbf{k_1},\mathbf{k_2},\mathbf{k_3})$ as $B_{\rm 21cm} (\mathbf{k_1},\mathbf{k_2},\mathbf{k_3}) = k_2^3 k_3^3/ (2 \pi^2)^2 b_{\rm 21cm} (\mathbf{k_1},\mathbf{k_2},\mathbf{k_3})$.

As done in Figure~\ref{fig:powerspectra}, we focus on wave numbers between $0.15 \leq k \leq 2 h~\rm cMpc^{-1}$ and show only two of the five F\&P averages for the sake of visual clarity. Further, we consider only two reference cases: 
\begin{itemize}
    \item \textbf{Equilateral triangles} ($B_{\rm equi}$): Here we set $k_1 = k_2 = k_3 = k$, where $k$ goes from 0.15 to $2~h~\rm cMpc^{-1}$. This allows us to explore the non-Gaussian features of the signal across various physical scales. The results are shown in Figure~\ref{fig:equi_bispectra}, with the same colours and linestyles as Figure~\ref{fig:powerspectra}. We note that the F\&P averages are a close match to the ensemble average across all redshifts. 
    The apparent large deviations seen at some wave modes of the F\&P average in comparison to the ensemble average (lower panels of Figure~\ref{fig:equi_bispectra}) arise because at those scales the bispectra approach 0. The normalisation by the ensemble average thus exaggerates the small differences significantly. \citet{Giri_2023} carried out a similar analysis for $z = 9$ (corresponding to $\langle x_{\rm HI} \rangle = 0.8$ in their simulations), and found the F\&P averages to be a close match to the ensemble average for $k > 0.1~h~\rm cMpc^{-1}$. Thus our results are consistent with their conclusions. \revised{We also quantify the improvement on using the F\&P method for $B_{\rm equi}$ in Appendix~\ref{sec:bequi_improve}, using the methodology of Section~\ref{sec:improve}, considering all our F\&P pair averages.}
    \item \textbf{Isosceles triangles} ($B_{\rm isoc}$): We set $k_1 = k_2 = k = 0.2~h~\rm cMpc^{-1} \neq k_3$ to explore large physical scales more thoroughly. We plot $B_{\rm isoc}$ versus the opening angle between the vectors $\mathbf{k}_1$ and $\mathbf{k}_2$ given as $\theta = \mathrm{cos}^{-1} (\mathbf{k}_1 \cdot \mathbf{k}_2 /(k_1 k_2))$ in Figure~\ref{fig:isoc_bispectra}, with the same colours and linestyles as Figure~\ref{fig:powerspectra}. As expected, the GIC simulations show large sample variance, which reduces with decreasing redshift. However, the F\&P averages are an even closer match to the ensemble average as compared to the GIC simulations, and thus continue to provide an improvement across all four of the redshift bins used. 
    Similarly to what observed for $B_{\rm equi}$, the large deviations seen at some scales at $z = 10$ are due to $B_{\rm isoc}$ approaching 0.
\end{itemize} 

We also repeat the process for other $B_{\rm isoc}$, by varying $k_1 = k_2 = k$ from 0.15 to 1.5 $h~\rm cMpc^{-1}$, and find that the trends hold across all physical scales. This is an interesting result, as it provides a useful statistic for modelling the 21-cm signal using F\&P averages at redshifts with $\langle x_{\rm HI} \rangle \leq 0.9$. It thus works well for comparing with observations too, as at these redshifts the bispectrum is a more useful statistic than the power spectrum.

However, it is necessary to verify that the improvement in estimating the ``true" bispectrum when using an F\&P average indeed correlates with statistical estimates of the physical properties (i.e., the distribution of neutral hydrogen) at the observed redshifts. A good way to check this, is to analyse the properties and distribution of the source of the non-Gaussian features of the $\delta T_{\rm b}$, i.e., the ionized regions.

\subsection{Bubble size distribution}\label{sec:bubble}

\begin{figure*}
\centering
\includegraphics[width=2.0\columnwidth,keepaspectratio]{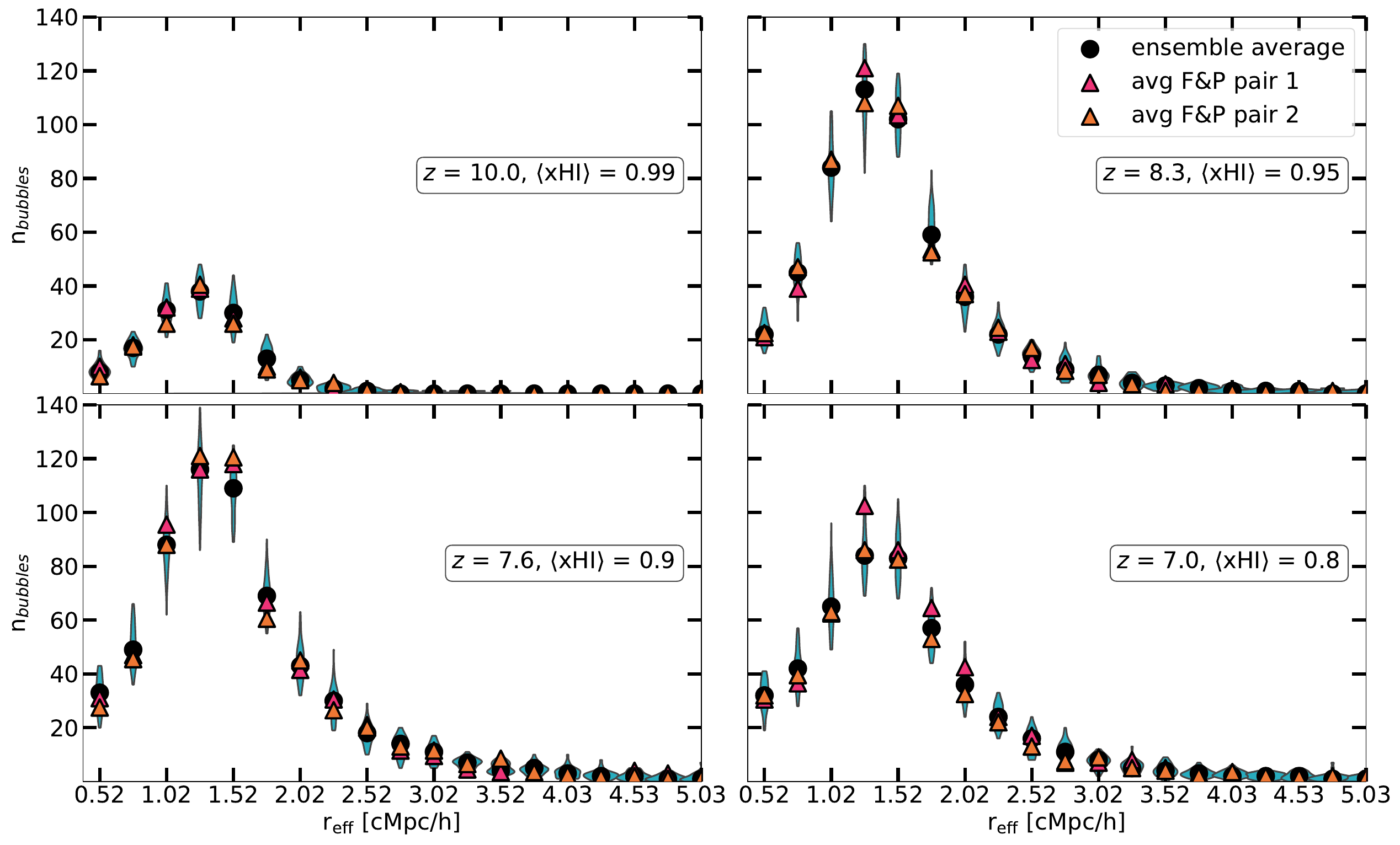}
\caption{Number of ionized bubbles in different radii $r_{\rm eff}$ bins at $z = 10, 8.3, 7.6,~\rm and~7$. We plot their distribution for the GIC simulations (cyan violin), the ensemble average (black circles), and the two F\&P averages used in Sections~\ref{sec:ps} and~\ref{sec:bs} (magenta and orange triangles).
} 
\label{fig:effradiiviolin}
\end{figure*}

There is no universal consensus on how to identify ionized regions, with several methods having been used in literature \citep[see][for a detailed comparison of different methods]{Giri_2018}. Here, firstly, we choose to define cells with $x_{\rm HI} \lesssim 0.5$ as ionized. Next, we use a Friend-of-Friends algorithm based on the {\tt ndimage} package of SciPy \citep{SciPy_2020} to identify regions with clusters of such ionized cells, which are then considered as ``bubbles''. From their volume, we derive the radius of an equivalent sphere and use it as an effective bubble radius ($r_{\rm eff}$). The threshold value of $x_{\rm HI}$=0.5 chosen to identify a cell as neutral or ionized is arbitrary. However, we find that varying this value does not affect our qualitative results.

The maximum value of $r_{\rm eff}$ is determined by the box size, while we ignore bubbles equivalent to an individual cell as they are resolution limited. We then find radii in the range $0.16 \lesssim r_{\rm eff} / (h^{-1}~\rm cMpc) \lesssim 40$. As our simulations are run only down to $z=7$ and quasars are not included as sources of ionizing photons, we do not expect many bubbles with $r_{\rm eff} > 5~h^{-1}~\rm cMpc$, and thus do not need to worry about the other extreme of the resolution range. We thus limit our analysis to the range $0.3 \leq r_{\rm eff} / (h^{-1}~\rm cMpc) \leq 5$, and build histograms of number of bubbles ($n_{\rm bubbles}$) binned according to $r_{\rm eff}$. In Figure~\ref{fig:effradiiviolin}, we construct violin plots (cyan) for the different values of $n_{\rm bubbles}$ for the GIC simulations, and also show the ensemble average (black circles). To compare this with the two F\&P averages used in Section~\ref{sec:bs}, we plot the averages for the number of bubbles for the individual fixed and paired simulations using the same colour scheme (magenta and orange triangles). Again, we only show two out of the generated five averages for the sake of easy visual comparison, as all five give similar results. We note that both F\&P averages are a good match for the ensemble average, and even when they deviate, they remain well within the violins, showcasing the ranges of the GIC simulations. This confirms that the improvement in the bispectrum noted when using the F\&P average is obtained because they closely match the number and sizes of ionized regions of the ensemble average.

Further, as expected, we see that the violin plots get narrower as we approach lower redshifts, as small scale variability between the GIC simulations grows with decreasing redshift. Lastly, we note that the number of smaller bubbles grows when going from $z = 10$ to $z = 7.6$, but by $z = 7$, many of them would have begun merging, leading to a fall in the number of bubbles with $1.0 \leq r_{\rm eff} / (h^{-1}~\rm cMpc)\leq 2.0$.

\section{Discussion}\label{sec:discuss}

\subsection{Advantage of the F\&P method}\label{sec:improve}

\begin{figure*}
\centering
\includegraphics[width=2.0\columnwidth,keepaspectratio]{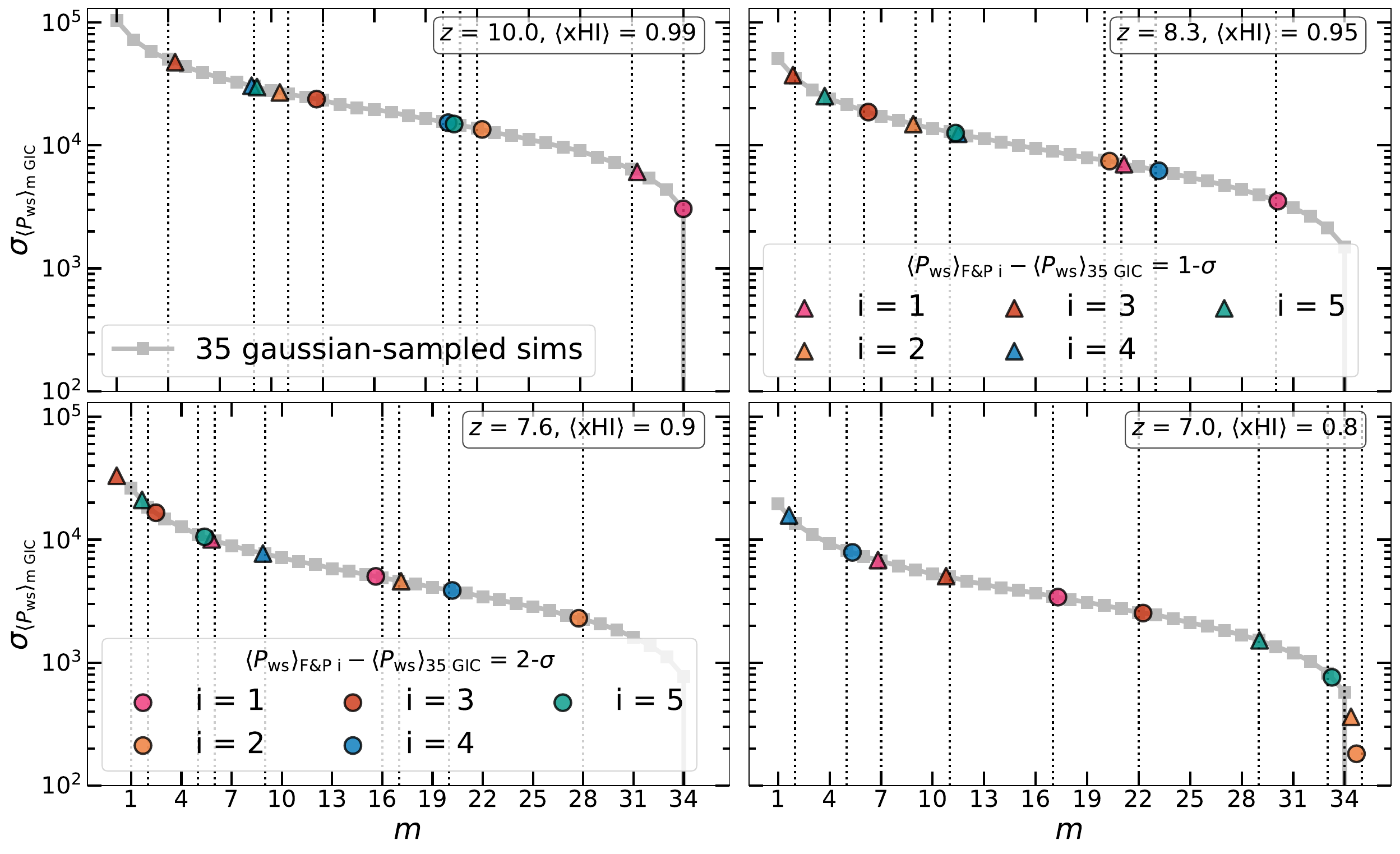}
\caption{
The standard deviation curve (grey) generated by interpolating the standard deviations of $\langle P_{\rm ws} \rangle_\mathrm{m~GIC}$ (grey squares) versus number of sampled GIC simulations $m$ going from 1 to 35 for $z = 10, 8.3, 7.6~\rm and~7$, clockwise from top left. The five F\&P averages are plotted as triangles (circles) with the assumption of them being 1-$\sigma$ (2-$\sigma$) away from $\langle P_{\rm ws} \rangle_\mathrm{35~GIC}$ using the same colours as used in Figures~\ref{fig:xHI_vs_z} and~\ref{fig:skew}. Their closest matches in the standard deviation curve are shown with black dotted vertical lines gives $m_{\rm eq}$.
} 
\label{fig:ps_improve}
\end{figure*}

Qualitatively, we note that the F\&P average provides a closer estimate of the ensemble average, as compared to any individual GIC simulation, for statistics like the skewness and the power spectrum, at least for $\langle x_{\rm HI} \rangle \geq 0.9$. For $\langle x_{\rm HI} \rangle \lesssim 0.9$, the improvements for the power spectrum is reduced and the bispectrum becomes the better statistic to use. 

However, it is necessary to quantify this improvement as compared to the average of multiple GIC simulations. For this reason, here we lay out the methodology to find the number of GIC simulations needed to match the performance of one F\&P average with respect to the power spectrum.
For this, we define ``performance'' as the extent of deviation of the average of multiple GIC simulations or of an F\&P pair from the power spectrum of the ensemble average. If the number of GIC simulations required to match the performance of an F\&P average is greater than 2, this means more of them need to be run to achieve the same performance. In this case, using the F\&P average which just needs 2 simulations to be run would reduce computational costs.

Ideally, for this comparison one should run a large number of GIC simulations as well as F\&P averages, and compare the extent of their deviation from the ensemble average at specific wave-modes. However, as radiation-hydrodynamical simulations are computationally expensive, we utilise the 35 GIC simulations discussed in previous sections, and compare them to all 5 F\&P averages generated in Section~\ref{sec:fandp}. To mimic large number statistics, we proceeds as follows:
\begin{enumerate}
    \item \textbf{Wave-mode window:} The availability of a large number of simulations allows the investigation of specific wave-modes, while the same cannot be done when we are limited to just a small number of them as in this case individual wave-modes are affected as well by random noise. We thus choose a window of wave-modes and sum all the power spectra at $k$s contained within it for each GIC simulation ($P_{\rm ws}$), their ensemble average ($\langle P_{\rm ws} \rangle_\mathrm{35~GIC}$), as well as each F\&P average ($\langle P_{\rm ws} \rangle_\mathrm{F\&P~i}$, where $i$ goes from 1 to 5). We choose this window to be $0.15 \leq k / (h~\rm cMpc^{-1}) \leq 0.4$ to ensure that the largest physical scales covered correspond to those at which the current and next-generation radio telescopes are most sensitive.
    \item \textbf{Choosing GIC simulations:} To evaluate the `effective volume' of the \revised{F\&P} simulations, we need to determine the number $m_{\rm eq}$ of GIC ones that match the statistical power of a single pair of \revised{F\&P} simulations. However, different subsets of $m < 35$ GIC simulations will produce different results, especially for small values of $m$. Therefore, in the following we consider all the $^{35} C_m$ possible subsets, which is the number of combinations of $m$ objects out of 35. But for computational reasons, we cap the number of combinations to $10^4$.
    \item \textbf{Standard deviation curve:} For every $m$, we compute the average $P_{\rm ws}$ for all $^{35} C_m$ combinations. These $\langle P_{\rm ws} \rangle_\mathrm{m~GIC}$ values form a Gaussian distribution centred around the ensemble average value $\langle P_{\rm ws}\rangle_\mathrm{35~GIC}$. We measure the standard deviation of this Gaussian distribution ($\sigma_{\langle P_{\rm ws} \rangle_\mathrm{m~GIC}}$) for each $m$. We linearly interpolate these values to generate the grey curves in Figure~\ref{fig:ps_improve}, with the grey squares referring to $\sigma_{\langle P_{\rm ws} \rangle_\mathrm{m~GIC}}$.
    \item \textbf{Standard deviation of the F\&P runs:} Although the above procedure should be repeated for the F\&P runs, this is prohibitively expensive from a computational point of view. Therefore, we explicitly parameterize our ignorance of the true width of the distribution of $\langle P_{\rm ws} \rangle_\mathrm{F\&P}$ by assuming that each F\&P average lies exactly 1-$\sigma$ away from its center. In other words, we assume that $\langle P_{\rm ws} \rangle_\mathrm{F\&P~i} - \langle P_{\rm ws} \rangle_\mathrm{35~GIC}$ is a measure of the width of the Gaussian distribution of $\langle P_{\rm ws} \rangle_\mathrm{F\&P}$. This is done independently for each F\&P run, and is shown in Figure~\ref{fig:ps_improve} with triangles. We then repeat this procedure but assuming that each F\&P average lies exactly 2-$\sigma$ away from the center of the distribution, and show the results with circles. 
    \item \textbf{Closest $m$ matching:} For the F\&P runs, we compare their standard deviation with the curve generated from the GIC simulations in step (iii) and determine the closest $m$ value (black dotted vertical lines in Figure~\ref{fig:ps_improve}). This is our estimate of $m_{\rm eq}$ for each F\&P simulation pair.
    \item \textbf{Improvement factor ($f_{\rm imp}$):} Finally, we define the improvement factor $f_{\rm imp} = m_{\rm eq}/2$ for all 5 F\&P runs. This corresponds to the ratio between the number of simulation runs (i.e. simulated volumes, since all our runs have the same box size) necessary with GIC and F\&P ICs, for which we report the minimum, maximum and average values in Table~\ref{table:ps_improve}.
\end{enumerate}

\begin{figure}
\centering
\includegraphics[width=\columnwidth,keepaspectratio]{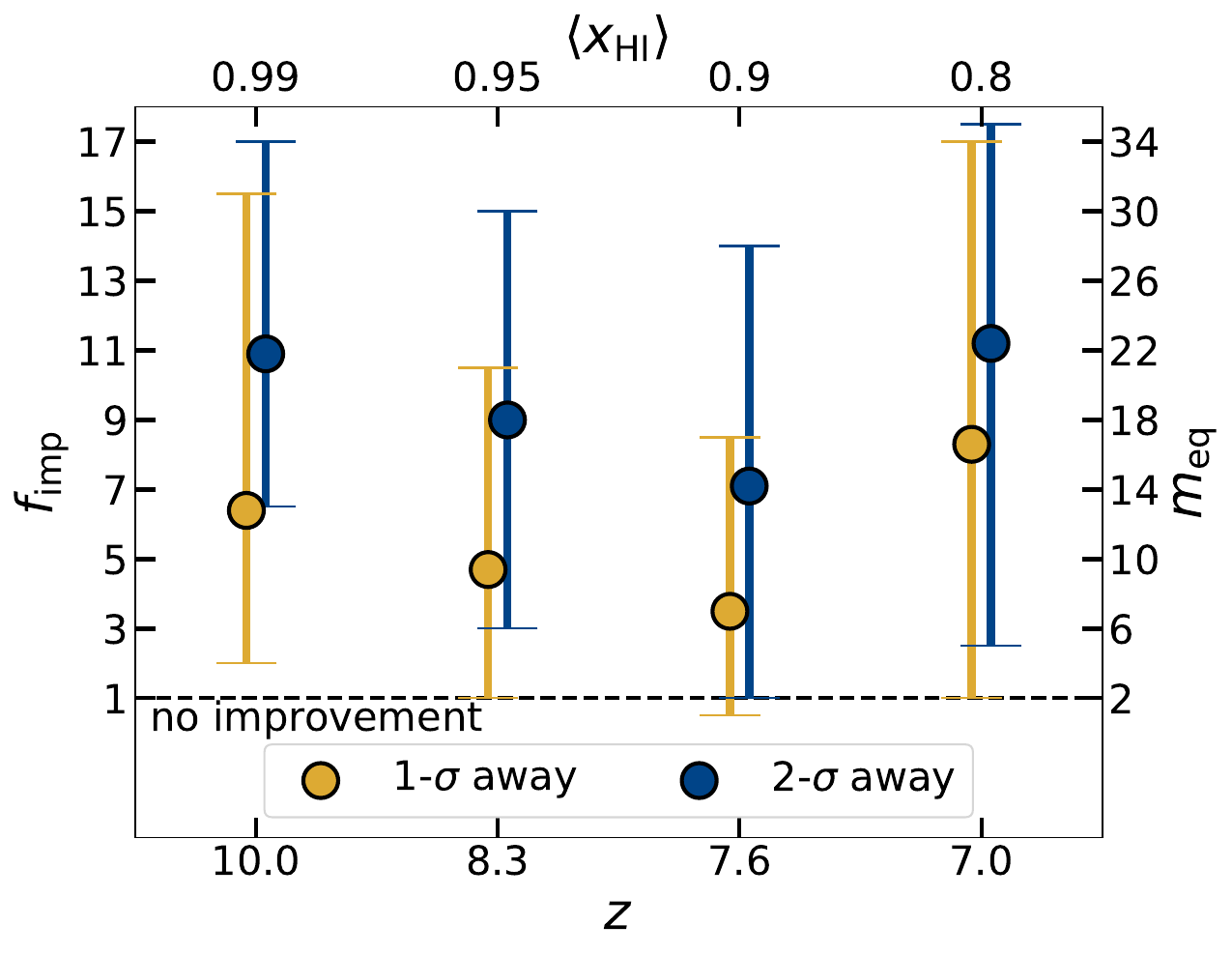}
\caption{Redshift evolution of the improvement factor in computational expense, $f_{\rm imp}$, when using the F\&P approach. The errorbars show the range from $f_{\rm imp, min}$ to $f_{\rm imp, max}$ for the F\&P averages being 1-$\sigma$ away (yellow) or 2-$\sigma$ away (blue) from an assumed distribution of F\&P averages, and the points show the average, $\langle f_{\rm imp} \rangle$. The top axis shows the corresponding $\langle x_{\rm HI} \rangle$ and the right axis shows the number $m_{\rm eq}$ of GIC simulations corresponding to $f_{\rm imp}$. The black dashed line indicates no improvement.
} 
\label{fig:fimp_vs_z}
\end{figure}

From Figure~\ref{fig:fimp_vs_z}, we note that the extent of improvement provided by the F\&P method reduces with decreasing redshift, as was expected from the qualitative results of Section~\ref{sec:ps}. However, interestingly, this trend seems to stop at $z = 7.6$, with higher values of $f_{\rm imp}$ noted at $z = 7$. This indicates that while the F\&P average may be performing worse at lower redshifts, it still does better than an average of a few GIC simulations. In fact, we note that $\langle f_{\rm imp} \rangle$ is $\sim 6$ at $z = 10$ (thus one F\&P average is better than running 12 GIC simulations), but $\sim 8$ at $z = 7$ (equivalent to 16 GIC simulations) for the case of F\&P averages being 1-$\sigma$ away. The lowest average improvement is 3.5 at $z = 7.6$, indicating that at least 7 GIC simulations are needed to match an F\&P average.

As we sum the power spectra across the aforementioned wave-mode window, a direct comparison between our results and those of \citet{Giri_2023} is difficult. Nevertheless, we note that in the range $10 \geq z \geq 7$, we obtain $f_{\rm imp} \geq 3.5$, which agrees with their result of an improvement of at least a factor of 4 at $k = 0.1~h~\rm cMpc^{-1}$ at $z = 9,~\langle x_{\rm HI}\rangle = 0.8$. The worst possible value of $f_{\rm imp}$ is 0.5, which corresponds to the case when we have a single GIC simulation perform better. This is expected, as pure randomness does allow such chance events. However, as the average improvement is above 2, we believe this still supports the use of F\&P averages for modelling the 21-cm signal power spectrum rather than running GIC simulations. 

\revised{A similar analysis can be run for the equilateral triangles bispectrum. We present this in more details in Appendix~\ref{sec:bequi_improve}, where we find $f_{\rm imp} \geq 5.0$. This result showcases that the F\&P method is even better for the bispectrum, and thus using multiple summary statistics for the 21-cm signal can allow us to maximise the interpretation of the 21-cm signal without requiring large effective volumes. Note that similar analyses can be performed for bispectra generated for different values of $k_1$, $k_2$ and $k_3$, by using a window over the opening angle instead of the wave-mode.}

\begin{table}
\centering
\caption{
$f_{\rm imp}$ is the factor of improvement on running an F\&P average over running multiple GIC simulations. We report the minimum, maximum and average value of this quantity for the two cases discussed in (iv), for the five F\&P averages.
}
\begin{tabular}{lllllll}
\hline 
$z$ & \multicolumn{3}{c}{$1-\sigma$ away} & \multicolumn{3}{c}{$2-\sigma$ away} \\
 & $f_{\rm imp, min}$ & $f_{\rm imp, max}$ & $\langle f_{\rm imp} \rangle$ & $f_{\rm imp, min}$ & $f_{\rm imp, max}$ & $\langle f_{\rm imp} \rangle$  \\
\hline
10.0 & 2.0 & 15.5 & 6.4 & 6.5 & 17.0 & 10.9 \\
8.3  & 1.0 & 10.5 & 4.7 & 3.0 & 15.0 & 9.0 \\
7.6  & 0.5 & 8.5 & 3.5 & 1.0 & 14.0 & 7.1 \\
7.0  & 1.0 & 17.0 & 8.3 & 2.5 & 17.5 & 11.2 \\
\hline
\end{tabular}
\label{table:ps_improve}
\end{table}

\subsection{Limitations and future applications}\label{sec:limits}

As we have shown in the previous Section, the F\&P method can be a powerful tool to extend the statistical accuracy of limited-volume simulations of the EoR 21-cm signal. 
In combination with the advancement of computational techniques and hardware, this method can deliver accurate predictions over volumes of interest for the study of the 21-cm signal, especially in the early phases of cosmic reionization. 
However, we caution that the F\&P method does not improve our predictive ability concerning individual galaxy properties, as they are dominated by local processes and environment. Therefore, the improvement granted on statistical quantities (e.g. the power spectrum and bispectrum) does not necessarily mirror exactly on different observables that might be more affected by individual objects (see e.g. the discussion in Sec.~\ref{sec:skew}).

We foresee numerous application of this technique. For instance, the lower computational cost required (with respect to traditional approaches) entails e.g. that larger volumes, as well as a broader range of cosmologies and/or physical models can be explored with accurate simulations, greatly improving the reliability of predictions and --~eventually~-- inference from 21-cm signal data. While basing the entire inference process on RMHD simulations remains prohibitive, they will be needed in order to confirm and improve constraints obtained through computationally-cheaper less-accurate methods. Additionally, they are necessary to explore the coupling between small and large scales, e.g. spatial correlations between galaxies and the 21-cm signal, that the advent of SKA will enable. Our results suggest that such accurate predictions can be obtained from RMHD simulations even in statistically-significant volumes of the Universe.

Finally, the F\&P approach operates orthogonally to super-resolution techniques \revised{\citep[][as discussed in Section~\ref{sec:intro}]{Kodi_2020,Li_2021}}; therefore these two approaches should be considered complementary rather than in opposition. 

\section{Summary}\label{sec:summary}

Running simulations for EoR is computationally very expensive. This is exacerbated when aiming at resolving low-mass galaxies \citep[whose importance has been recently shown observationally in][]{Atek_2023} in volumes large enough to be statistically significant for reionization studies. In this work, we explored the Fixed \& Paired (F\&P) approach \citep[][]{Angulo_2016,Pontzen_2016} to investigate the possibility of reducing the number of simulations needed (and thus the overall computational expense) to produce unbiased models of the 21-cm signal. While past efforts have used semi-numerical approaches to implement this, we have shown more rigorous results by using radiation hydrodynamic simulations that model more accurately galaxy-scale effects on the 21-cm signal. We focus on the wave modes in the range $0.15 \leq k / (h~\rm cMpc^{-1}) \leq 2 $, as the best measurements from present and upcoming radio telescopes like LOFAR, HERA, MWA and SKA are expected in this regime. Further, we focus on redshifts $10 \geq z \geq 7$, which in our case correspond to $1.0 > \langle x_{\rm HI} \rangle \geq 0.8$.

To explore the improvement with respect to various 21-cm signal statistics obtained by adopting the F\&P approach rather than running Gaussian-sampled initial conditions based (GIC) simulations, we use a setup similar to that of the {\tt THESAN} project \citep[][]{Kannan_2022, Garaldi_2022,Smith_2022,Garaldi_2023}. In particular, we investigate the impact on the skewness, power spectrum, bispectrum and the ionized region size distribution, and introduce a novel method to quantify the improvement.  

We find that the skewness and power spectrum are well-estimated by the F\&P averages for $\langle x_{\rm HI} \rangle \geq 0.9$, and their performance is good also in the range $0.9 > \langle x_{\rm HI} \rangle \geq 0.8$, with an improvement in computational cost better than 3.5 for the generation of the power spectrum. 
We find that the bispectrum  is well estimated for $\langle x_{\rm HI} \rangle \geq 0.8$, with the scaled deviation between the F\&P averages and the ensemble average being below $1$. The only exception are those modes where the ensemble average bispectrum approaches zero, artificially increasing the small differences between the F\&P average and the ensemble average. \revised{In fact, we find that the improvement in computational cost is better than a factor of 5 for the equilateral triangle bispectrum.} This confirms that the F\&P average bispectrum is a great complement to the power spectrum for studies of the 21-cm signal when $\langle x_{\rm HI} \rangle \geq 0.8$. Finally, we show that in this regime the F\&P averages also provide a good estimate of the HII regions size distribution, with the F\&P averages being within $2-\sigma$ deviation of the ensemble average for bubbles with radius $\leq 5 h^{-1}$ cMpc.

Thus, the F\&P method can be used to model the 21-cm signal summary statistics with significantly reduced effective volume and computational expense.

\section*{Acknowledgements}
The authors would like to thank Sambit K. Giri and Raul Angulo for useful comments and discussions. \revised{The authors would also like to thank the referee for their helpful comments that helped improve this paper significantly.}

\section*{Data Availability}
The data used in this work can be shared upon reasonable request. 

\bibliographystyle{mnras}
\bibliography{fandp_thesan} 

\appendix 
\section{Improvement factor for the equilateral bispectrum}\label{sec:bequi_improve}

\begin{figure*}
\centering
\includegraphics[width=2\columnwidth,keepaspectratio]{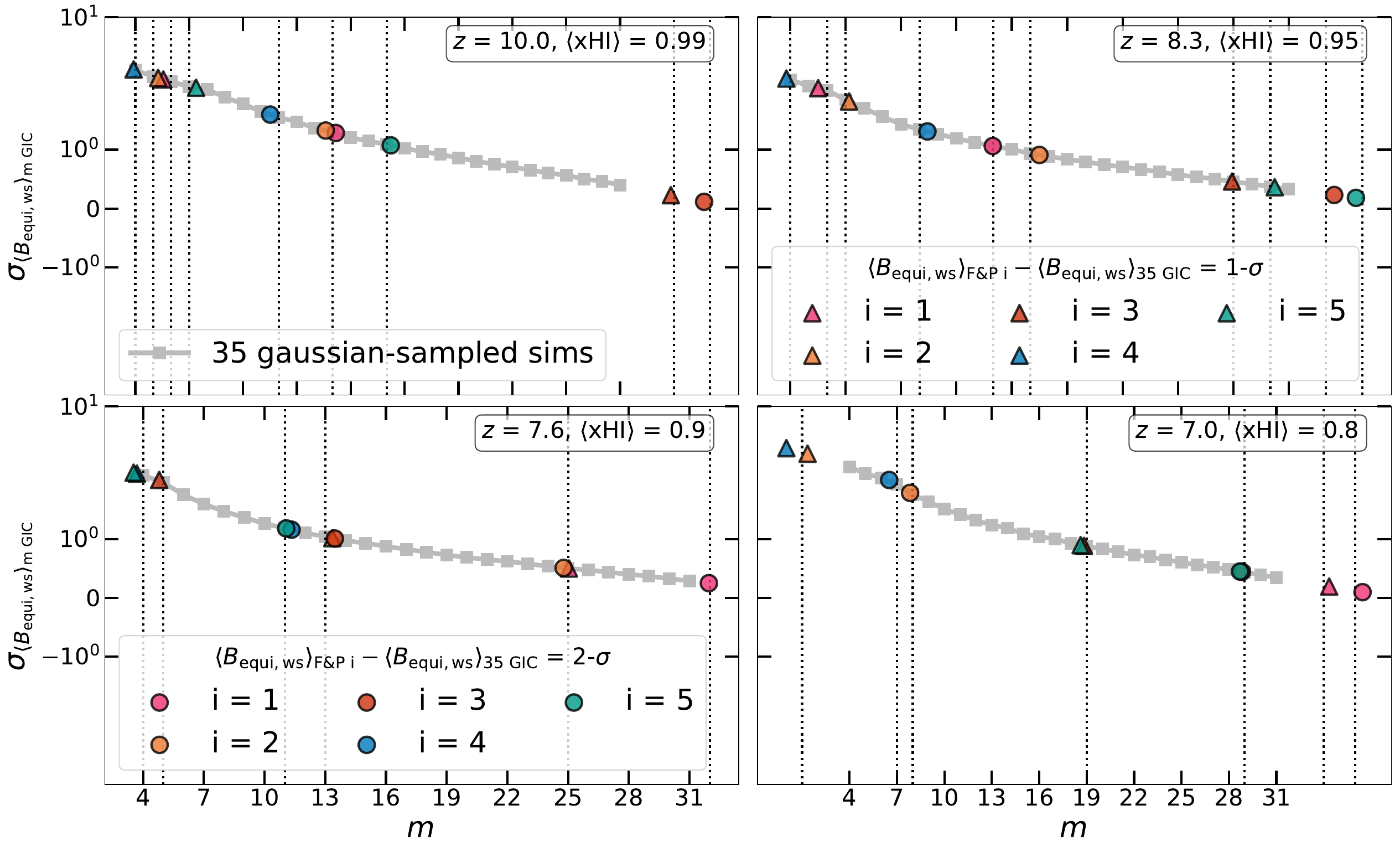}
\caption{ As Figure~\ref{fig:ps_improve} for the equilateral triangle bispectrum ($B_{\rm equi}$).
} 
\label{fig:Bequi_improve}
\end{figure*}

\begin{figure*}
\centering
\includegraphics[width=\columnwidth,keepaspectratio]{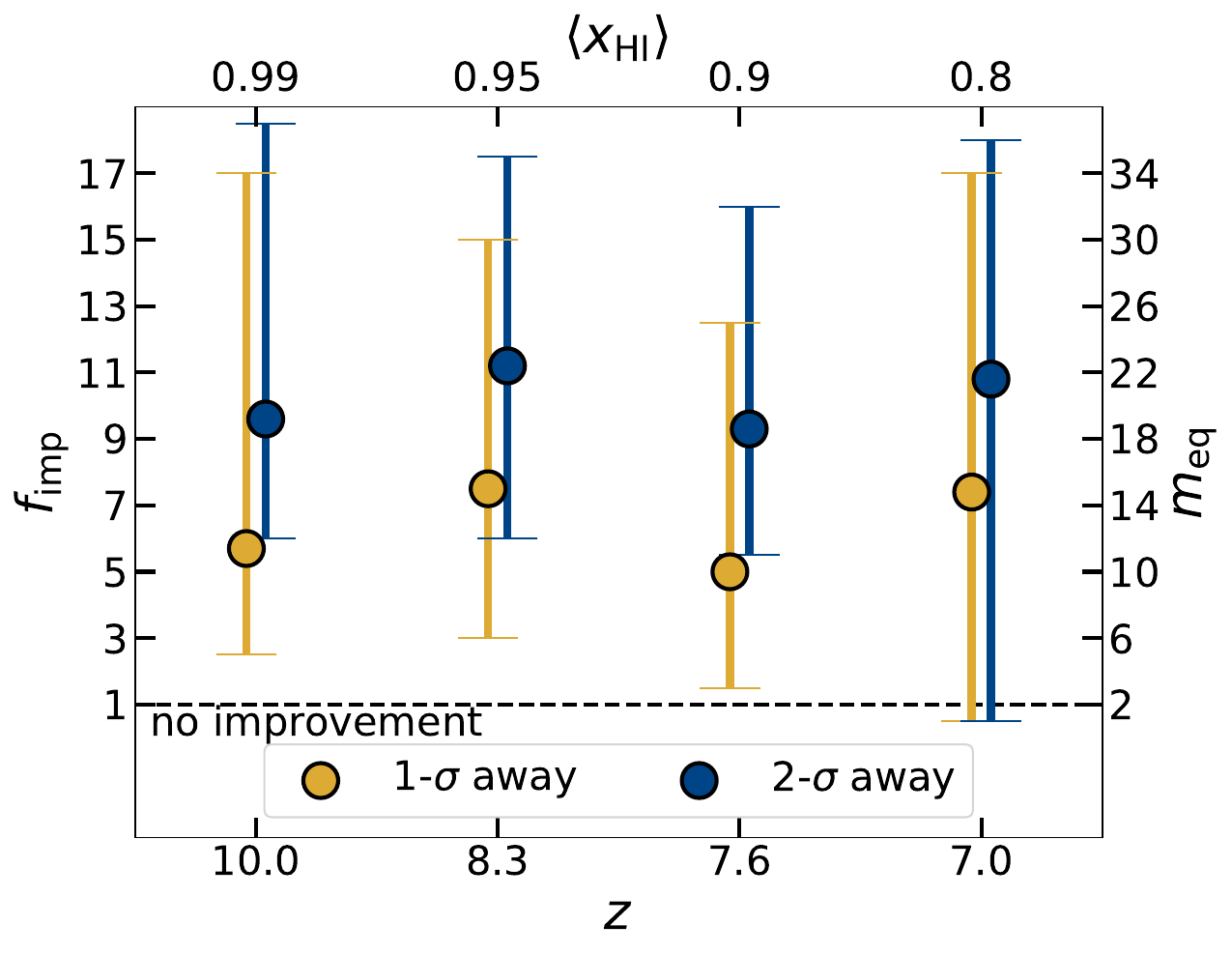}
\caption{As Figure~\ref{fig:fimp_vs_z} for the equilateral triangle bispectrum ($B_{\rm equi}$).
} 
\label{fig:Beq_fimp_vs_z}
\end{figure*}

\revised{We repeat the analysis of the improvement factor when using the F\&P method as done in Section~\ref{sec:improve} for the bispectrum. While this can be performed for any type of bispectra, for the sake of simplicity here we discuss only  $B_{\rm equi}$. For a direct comparison with the power spectrum improvement factor, we use the same wave-mode window of $0.15 \leq k / (h~\rm cMpc^{-1}) \leq 0.4$.

Further, for the standard deviation curve for $B_{\rm equi}$ shown in Figure~\ref{fig:Bequi_improve}, we do not consider combinations of GIC simulations with $m < 4$ or $m > 32$ because of sampling bias, which is significant because unlike the power spectrum, the equilateral bispectrum can take negative values. This, combined with the fact that the values of $B_{\rm equi}$ are close to 0 mean that minor deviations from the ensemble average can lead to severe deviations in the average in cases of poor sampling. As done in Section~\ref{sec:improve}, we report the minimum, maximum and average improvement factor ($f_{\rm imp}$) values in Table~\ref{table:Bequi_improve} and depict the same graphically in Figure~\ref{fig:Beq_fimp_vs_z}. We note that the average improvement factor for $B_{\rm equi}$ is consistent across redshifts $10 \geq z \geq 7$.}

\begin{table*}
\centering
\caption{
As Table~\ref{table:ps_improve}, showing the $f_{\rm imp}$ improvement factor for the equilateral triangle bispectrum ($B_{\rm equi}$).
}
\begin{tabular}{lllllll}
\hline 
$z$ & \multicolumn{3}{c}{$1-\sigma$ away} & \multicolumn{3}{c}{$2-\sigma$ away} \\
 & $f_{\rm imp, min}$ & $f_{\rm imp, max}$ & $\langle f_{\rm imp} \rangle$ & $f_{\rm imp, min}$ & $f_{\rm imp, max}$ & $\langle f_{\rm imp} \rangle$  \\
\hline
10.0 & 2.5 & 17.0 & 5.7 & 6.0 & 18.5 & 9.6 \\
8.3  & 2.0 & 15.0 & 7.5 & 6.0 & 17.5 & 11.2 \\
7.6  & 2.5 & 12.5 & 5.0 & 5.5 & 16.0 & 9.3 \\
7.0  & 0.5 & 17.0 & 7.4 & 3.0 & 18.0 & 10.8 \\
\hline
\end{tabular}
\label{table:Bequi_improve}
\end{table*}

\bsp	
\label{lastpage}
\end{document}